\documentclass[
reprint,
longbibliography,
showkeys,
amsmath,amssymb,
aps,
pre, 
onecolumn
]{revtex4-2}

\usepackage{graphicx}
\usepackage[justification=justified]{subcaption}

\usepackage{amsmath}
\usepackage{amssymb}
\usepackage{mathtools}
\usepackage{ bbold }
\usepackage{float}
\usepackage{braket}
\usepackage{algorithm}
\usepackage{algpseudocode}     
\usepackage{dcolumn}
\usepackage{bm}

\usepackage{hyperref}

\usepackage[mathscr]{euscript}   

\usepackage{tcolorbox}
\usepackage{comment}
\usepackage{float}

\newcommand{\er}[1]{Eq.\,\eqref{#1}}

\newcommand{\era}[2]{Eqs.\,(\ref{#1}) and (\ref{#2})}

\newcommand{\Er}[1]{Equation~\eqref{#1}}

\usepackage{hyperref}
\hypersetup{
colorlinks=true,
linkcolor=blue,
filecolor=blue,
citecolor=blue,      
urlcolor=blue,
}
\usepackage{color}

\makeatletter
\newsavebox{\@brx}
\newcommand{\llangle}[1][]{\savebox{\@brx}{\(\m@th{#1\langle}\)}%
  \mathopen{\copy\@brx\kern-0.5\wd\@brx\usebox{\@brx}}}
\newcommand{\rrangle}[1][]{\savebox{\@brx}{\(\m@th{#1\rangle}\)}%
  \mathclose{\copy\@brx\kern-0.5\wd\@brx\usebox{\@brx}}}
\makeatother

\begin{document} 


\title{Self-interacting processes via Doob conditioning}

\author{Francesco Coghi}
\thanks{\textit{francesco.coghi@nottingham.ac.uk}}

\author{Juan P.\ Garrahan}
\affiliation{School of Physics and Astronomy, University of Nottingham, Nottingham, NG7 2RD, United Kingdom}
\affiliation{Centre for the Mathematics and Theoretical Physics of Quantum Non-Equilibrium Systems, University of Nottingham, Nottingham, NG7 2RD, United Kingdom}

\begin{abstract}
We connect self-interacting processes, that is, stochastic processes where transitions depend on the time spent by a trajectory in each configuration, to Doob conditioning. In this way we demonstrate that Markov processes with constrained occupation measures are realised optimally by self-interacting dynamics. 
We use a tensor network framework to guide our derivations. We illustrate our general results with new perspectives on well-known examples of self-interacting processes, such as random walk bridges, excursions, and forced excursions. 
\end{abstract}

\maketitle

\section{Introduction}

A self-interacting process (SIP) is a stochastic process whose transition probabilities at any time depend on the {\em empirical occupation} of configurations up to that time (i.e., on the fraction of time spent by the current trajectory in the configurations of the system under consideration). In this way, the behaviour of the trajectories of a SIP is shaped by how frequently it has visited past configurations. This type of dynamics extends beyond the traditional framework of Markov processes, aiming to capture more complex, memory-dependent systems that frequently arise in real-world contexts. For instance, the dynamics of autochemotactic processes~\cite{budrene1995dynamics,tsori2004self-trapping,reid2012slime,zhao2013psl-trails,kranz2016effective,meyer2021optimal} (self-guided entities like ants navigating towards food in response to their own chemical secretions) can be modelled by SIPs. 
Alternatively, while the underlying microscopic dynamics may be fully Markovian, under coarse-graining, or due to measurement limitations or incomplete data, memory effects can emerge~\cite{zwanzig2001nonequilibrium,espanol2009mori--swanzig,li2015incorporation,pasquale2019systematic,brandner2024dynamics}, giving rise to effective dynamics that may be described by a SIP. 

The study of SIPs, particularly their long-time behaviour, has been an active area of research in probability. Notable examples discussed in the survey~\cite{pemantle2007a-survey} include reinforced random walks (on edges or nodes), self-avoiding walks, and self-interacting Markov chains. Self-avoiding walks~\cite{toth2001self-interacting} have practical applications in polymer physics, while self-interacting Markov chains have been used to model population dynamics in biology~\cite{pemantle2007a-survey}, and are highly valuable in improving the performance of Monte Carlo methods for sampling quasi-stationary distributions~\cite{moral2007self-interacting,kannan2008self-interacting,budhiraja2022empirical}. These models have also been extended to continuous-time dynamics, which in the Brownian case are referred to as {\em self-interacting diffusions}. This extension, explored in Refs.\,\cite{benaim2002self-interacting,bottou2003stochastic,benaim2005self-interacting,kurtzmann2010the-ode-method,benaim2011self-interacting,chambeu2011some,kleptsyn2012ergodicity}, builds on a dynamical systems framework known as stochastic approximation~\cite{benaim1999dynamics}. 

In this paper we show how non-Markovian SIPs can be understood in terms of the rare events of Markovian processes within the framework of Doob conditioning~\cite{doob1957conditional,doob1984classical,borkar2003peformance,jack2010large,chetrite2015nonequilibrium}. 
Specifically, we show that if a Markovian process whose trajectory measure is either conditioned or biased (viz.\ it is subject to either a hard or a soft condition) through its empirical occupation measure, then the optimal sampling dynamics (i.e., the {\em unconditioned} dynamics that most efficiently reproduces the condition, also known as the ``Doob'' dynamics \cite{chetrite2015nonequilibrium}) can be formulated as a SIP. We illustrate this perspective with simple examples where the unconditioned dynamics is that of a fair binary coin, meaning that the dynamics of the empirical occupation is that of an unbiased random walk (RW): using our general approach it is easy to see that well-studied problems like RW bridges and excursions~\cite{majumdar2005airy,majumdar2006brownian,majumdar2008time,majumdar2015effective,szavits2015inequivalence,causer2022slow,mazzolo2022conditioning,mazzolo2023joint} can be interpreted as instances of SIPs.

The paper is organised as follows. In Sec.\,\ref{sec:model} we establish our notation for the discrete time and discrete state settings that we consider, and define SIPs in this context. In Sec.\,\ref{sec:condmps} we introduce the trajectory observables used for both hard (cf.\ microcanonical) and soft (cf.\ canonical) trajectory conditioning. We also define the tensor network notation and visualisation that we use as a conceptually simple way to complement our analytical results on the connection between conditioning and non-Markovian dynamics. In Sec.\,\ref{sec:Doob} we derive our central result using Doob conditioning. Section \ref{sec:examples} presents three illustrative examples related to conditioned random walks. We conclude in Sec.\,\ref{sec:conclusion} with a summary and discussion.

\section{Self-interacting processes in discrete time}
\label{sec:model}

\subsection{Original unbiased dynamics}

We consider a discrete-time Markov chain \cite{grimmett2020probability} $(X_t)_{0 \leq t \leq T}$, where $X_t$ takes values in a finite, discrete state space $\Xi$ of cardinality $|\Xi\, | \, = d$. The probabilities of transitioning between any two states $x$ and $y$ are encapsulated in the transition probability matrix $\left\lbrace p(x,y) \right\rbrace_{x,y=1:d}$, with elements defined as the conditional probabilities
\begin{equation}
    p(x,y) \coloneqq \mathbb{P}[X_{t + 1} = y \, | \, X_t = x] \, ,
    \label{eq:pxy}
\end{equation}
such that $p(x,y) \geq 0$ for all $x,y \in \Xi$, with 
\begin{equation}
    \label{eq:norm}
    \sum_{y \in \Xi} p(x,y) = 1
    \, ,
\end{equation}
and where we use the symbol $\mathbb{P}[\cdot]$ to indicate ``probability of $(\cdot)$''. We assume for simplicity that the transition probabilities are independent of time (i.e., independent of the step $t$ in the chain). The initial state $X_0$ of a trajectory is drawn from an initial probability distribution, $X_0 \sim p_0(\cdot) \coloneqq \left\lbrace p_0(x) \right\rbrace_{x=1:d}$, with
\begin{equation}
\label{eq:InitialDistributtion}
  	p_0(x) \coloneqq \mathbb{P}[X_0 = x] \, . 
\end{equation}

We denote by $\Omega_T$ the sample space of the Markov chain, which we alternatively call {\em trajectory space}, for a (finite) time horizon $T$ (i.e., the overall time extent of trajectories). This is given by the $(T+1)$-fold tensor product of the state space, $\Omega_T = \Xi^{\otimes (T+1)}$. Each $\omega_{0:T} \in \Omega_T$ represents a possible sequence of $T$ states (together with the initial state) under the dynamics defined by \er{eq:pxy} up to time $T$. Using the Markov property, the probability of a trajectory factorises as 
\begin{equation}
\label{eq:PathProba}
    \mathbb{P}[\omega_{0:T} = (x_0, x_1, x_2, \cdots, x_T)]
        = 
    	\mathbb{P}[X_0 = x_0, X_1 = x_1, \cdots, X_T = x_T]
        = 
        p_0(x_0) \prod_{t=1}^{T} p(x_{t-1},x_t) \, .
\end{equation}

\subsection{Empirical occupation measure}

An important dynamical quantity in what follows will be the {\em empirical occupation measure} \cite{touchette2009the-large}. This refers to the trajectory-dependent (and time-dependent) probability of configurations as inferred from that trajectory at each time. It is computed from the number of visits to each configuration in the trajectory. Specifically, 
\begin{equation}
\label{eq:EmpiricalOccupation}
	\rho_t(x \, | \, \omega_{0:t}) \coloneqq \frac{1}{t+1} \sum_{t'=0}^t \delta_{x,X_{t'}} \, . 
\end{equation}
The empirical $\rho_t( \cdot \, | \, \omega_{0:t})$ takes values in $\Gamma$ which is a discrete subset of the simplex $[0,1]^d$ (filling the simplex only in the $t \to \infty$ limit). It is also by definition normalised, $\sum_{x \in \Xi} \rho_t( x \, | \, \omega_{0:t}) = 1$, for all $\omega_{0:t}$ and all $t$. The set of empiricals along a trajectory $\omega_{0:T}$ are related via the recursion
\begin{equation}
\label{eq:EmpOccDynamics}
  	(t+1) \rho_t( x \, | \, \omega_{0:t}) = t \rho_{t-1}( x \, | \, \omega_{0:t-1}) + \delta_{X_{t}, x} 
    \, ,
\end{equation}
where $\omega_{0:t}$ in the equation above is the portion of trajectory $\omega_{0:T}$ up to time $t$.

\subsection{Self-interacting processes}

We define a self-interacting chain $(\tilde{X}_t)_{0 \leq t \leq T}$ as a stochastic process, also with state space $\Xi$, but where the transition rates depend not only on the current and next state, as in a Markov chain, but also on the empirical occupation at that time. That is, if up to time $t$ the current trajectory is $\omega_{0:t}$, then the
transition probabilities at time $t+1$ have the general form
\begin{equation}
    \label{eq:TransitionSIMC}
    p_{t+1}(x,y) \coloneqq \mathbb{P}
    \left[
        \tilde{X}_{t + 1} = y \, | \, \tilde{X}_t = x, 
        \rho(\cdot \, | \, \omega_{0:t})
    \right] 
    \, .
\end{equation}
The function above is a mapping from the space $\Xi \times \Gamma$ to $\Xi$, in contrast to the Markovian \er{eq:pxy} which maps $\Xi$ to $\Xi$. With the definition \eqref{eq:TransitionSIMC}, $p_{t+1}(x,y)$ is a transition matrix that is indexed by the empirical occupation measure as in Ref.\,\cite{moral2007self-interacting}. 

It is important to note that while the sequence $(\tilde{X}_t)_{0 \leq t \leq T}$ is no longer a Markov chain, the joint process $(\tilde{X}_t, \rho_t)_{0 \leq t \leq T}$ can be described as a (in general time-inhomogeneous) Markov chain in the extended state space $\Xi \times \Gamma$. Such a property is key for deriving the main result of this paper, as well as for describing the dynamics of SIPs in tensor network language, as shown in the following section.

\section{Conditioning observables and tensor network framework}
\label{sec:condmps}

\subsection{Observables: hard and soft conditioning}

We consider again the original Markov chain of Sec.\,II.A. 
We introduce the trajectory observable
\begin{equation}
        \label{eq:Conditioning}
        \mathbf{F}_T(\omega_{0:T}) \coloneqq \left( f_1(\rho_1(\cdot \, | \, \omega_{0:1})), \cdots, f_T(\rho_T(\cdot \, | \, \omega_{0:T})) \right) \, ,
\end{equation}
where the functions $\{ f_t \}_{1:T}$, which may explicitly depend on time, map $\rho_t$ into $\mathbb{R}$.

\subsubsection{Hard conditioning}

In the case of {\em hard conditioning} we require the $T$-dimensional observable \er{eq:Conditioning} to satisfy a hard constraint. Given a $T$-dimensional set of segments on the real line, $\mathbf{D}_T = (D_1, \cdots, D_T)$, with $D_t \subset \mathbb{R}$, we impose the condition
\begin{equation}
    \label{eq:ConditioningValues}
    \mathbf{F}_T(\omega_{0:T}) \in \mathbf{D}_T 
    \, .
\end{equation}
The above imposes dynamical constraints on the process $(X_t)_{0 \leq t \leq T}$. The conditioned process must therefore evolve such that at each time step the empirical measure satisfies \eqref{eq:ConditioningValues}. In the language of statistical mechanics, for a trajectory ensemble this is analogous to working within a micro-canonical ensemble of configurations, where specific constraints must be strictly adhered to.

\subsubsection{Soft conditioning}

We may alternatively require the $T$-dimensional observable \eqref{eq:Conditioning} only satisfies \er{eq:ConditioningValues} approximately. We can do this by reweighting the probability of trajectories of the stochastic process such that $\mathbf{F}$ lies mostly within the desired set $\mathbf{D}_T$. We call this {\em soft conditioning}. The statistical mechanics analogy in this case is that of the canonical ensemble, where conditions are only enforced on average. 

We show in Sec.\,IV below that the optimal sampling dynamics of a Markov process for any combination of hard and/or soft conditioning \eqref{eq:ConditioningValues}, with $\mathbf{F}_T$ defined as in \er{eq:Conditioning}, is a self-interacting Markov chain. 


\begin{figure*}[t]
    \centering
    \includegraphics[width=\textwidth]{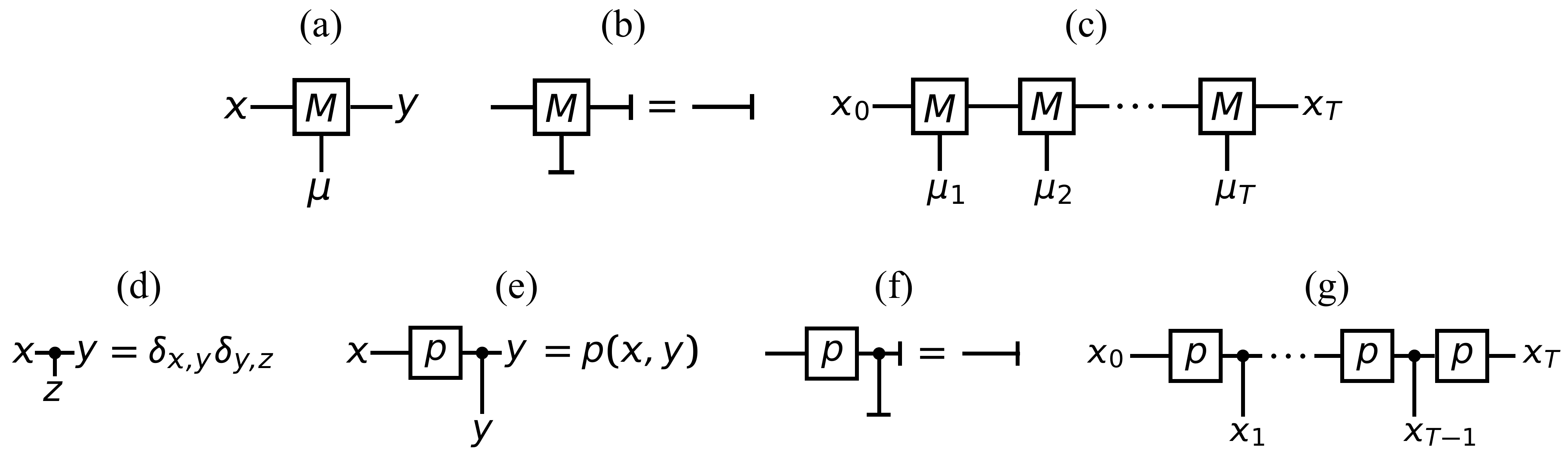}
    \caption{
        {\bf Tensor network representation of a hidden Markov process.} 
            {\bf (a)} Rank-3 tensor corresponding to the transition operator \er{eq:M}. The horizontal legs have dimension $\Xi$ (or ``bond dimension'') of the hidden states. The vertical leg has dimension $\Gamma$ of the signal. Time flows from left to right.
            {\bf (b)} Normalisation \er{eq:sumM}. The T-shaped operators represent the ``flat'' states in either $\Xi$ or $\Lambda$. The joining of legs indicate contraction over the indices. One calls this a TN in {\em right canonical form}.
            {\bf (c)} Tensor ``train'' encoding the trajectories of the process up to time $T$. By assigning specific values to the legs of the tensor one obtains the probability of that specific trajectory. 
        {\bf Tensor network representation of a classical Markov process.} 
            {\bf (d)} Rank-3 delta tensor, which vanishes unless all legs take the same value. 
            {\bf (e)} For a Markov process the signal is given by the states, and the elementary tensor of panel (a) reduces to the one shown corresponding to \er{eq:pxy}. 
            {\bf (f)} Normalisation condition (or right canonical form of the TN) given by \er{eq:norm}.
            {\bf (g)} TN for a trajectory of the Markov process. 
    }  
    \label{fig:fig1}
\end{figure*}

\subsection{Trajectory ensembles and tensor networks}

A convenient framework for discussing problems relating to ensembles of stochastic trajectories is that of tensor networks (TNs) \cite{verstraete2008matrix,schollwock2011the-density-matrix,orus2014a-practical,silvi2019the-tensor,okunishi2022developments,banuls2023tensor}, specifically matrix product states (MPSs) and matrix product operators (MPOs). While TN methods have been mostly used to study problems in quantum many-body systems \cite{verstraete2008matrix,schollwock2011the-density-matrix,orus2014a-practical,silvi2019the-tensor,okunishi2022developments,banuls2023tensor}, they are gaining acceptance in classical stochastic dynamics
\cite{gorissen2009density-matrix,gorissen2012current,garrahan2016classical,banuls2019using,helms2019dynamical,helms2020dynamical,causer2020dynamics,causer2022finite,strand2022using,causer2023optimal,zima2025chemical}. 

It will prove helpful to consider in general classical {\em hidden} Markov processes (of which classical Markov processes are a subclass) \cite{grimmett2020probability}. For these, the states of the system are not assumed to be observable, and the only accessible information in a stochastic trajectory, apart from the initial and final configurations, is a signal at each transition along the trajectory (which in general may not be enough to resolve the states before and after). The transition probabilities are encoded in a rank-3 tensor 
\begin{equation}
    \label{eq:M}
    \{ M_\mu(x,y) \}_{x,y \in \Xi, \mu \in \Lambda}
    \, ,
\end{equation}
where $\Lambda$ indicates the set of possible signals. This tensor can be represented graphically as in Fig.\,1(a). For a classical (hidden) Markov chain, the tensors $M$ have to be real and positive, and probability conservation requires that 
\begin{equation}
    \label{eq:sumM}
    \sum_{y \in \Xi} \sum_{\mu \in \Lambda} M_\mu(x,y) = 1 
    \; , \; \forall x \in \Xi
    \, . 
\end{equation}
The graphical representation of \er{eq:sumM} as a vectorial equation is shown in Fig.\,1(b): the r.h.s.\ is the unit or {\em flat} vector, and in the l.h.s.\ joined lines indicate contraction (i.e., sum over components). The ensemble of trajectories is the TN obtained by contracting the tensors \eqref{eq:M}: Fig.\,1(c) shows graphically the component of the TN corresponding to the probability of the trajectory starting at $x_0$, ending at $x_T$ and conditioned on the signals $(\mu_1, \cdots, \mu_T)$. This TN is called an MPS since once the values of the signals $\mu_t$ are fixed, the probability of the trajectory is obtained by multiplying the matrices $M_{\mu_t}(\cdot, \cdot)$. 
In TN jargon, the state space $\Xi$ is the {\em bond dimension} and the space $\Lambda$ of signals, the {\em physical dimension}. Since we are using TNs to represent {\em exactly} the stochastic dynamics, both $\Xi$ and $\Lambda$ are fixed by the problem under study (this is in contrast to the use of TNs as an approximate representation of functions, where the bond dimension is often used as a variational parameter). 

For classical Markov chains where the states are fully observable, the signal is simply given by the states involved in the transition. Using the delta tensor, Fig.\,1(d), the transition operator Fig.\,1(a) can be simplified to the form shown in Fig.\,1(e), its components given by the transition probabilities \eqref{eq:pxy}. The normalisation condition \eqref{eq:norm} is shown graphically in Fig.\,1(f). By contracting the elementary tensors of Fig.\,2(c), we obtain the TN that encodes the ensemble of trajectories, and whose components are the trajectory probabilities, see Fig.\,1(g).

\section{Generalised Doob conditioning}
\label{sec:Doob}

In this section we show how a Markov chain $(X_t)_{0 \leq t \leq T}$ (as defined in Sec.\,II.A), conditioned on particular values of the observable $\mathbf{F}_T(\omega_{0:T})$ as in \er{eq:ConditioningValues}, can be seen as a self-interacting Markov chain $(\tilde{X}_t)_{0 \leq t \leq T}$ as defined in Sec.\,II.C. In the hard-conditioning scenario, cf.\ Sec.\,III.A.1, we write
\begin{equation}
(X_t)_{0 \leq t \leq T}\Big|_{\mathbf{F}_T(\omega_{0:T}) \in \mathbf{D}_T} 
    = (\tilde{X}_t)_{0 \leq t \leq T} \, ,
\end{equation}
where the long bar $\Big|$ stands for ``conditioned to'', whereas in the soft-conditioning scenario we introduce the notation
\begin{equation}
(X_t)_{0 \leq t \leq T} \Big|_{W[\mathbf{F}_T(\omega_{0:T}), \mathbf{D}_T]} 
    = (\tilde{X}_t)_{0 \leq t \leq T} \, ,
\end{equation}
where $W[\mathbf{F}_T(\omega_{0:T}), \mathbf{D}_T]$ represents the weighting function applied to favour paths close to $\mathbf{D}_T$, with a weighting $W$ that will be specified later on. 

Conditioning also has a simple representation in terms of TNs. For example, the l.h.s.\ of Fig.\,2(a) corresponds to local-in-time conditioning (or biasing) the trajectory ensemble of Fig.\,1(c): the operator $W$ is diagonal in the signal, $(W_\mu \delta_{\mu, \nu})_{\mu,\nu \in \Lambda}$, so that for each transition $\mu$ in the trajectory its probability is reweigthed by a factor $W_\mu$.

\subsection{One time Doob conditioning}
\label{sec:DoobClassical}

Before addressing the general case, we first review Doob's classical result for conditioning at a fixed time~\cite{doob1957conditional,doob1984classical} (for a modern exposition of these ideas see Ref.~\cite{chetrite2015nonequilibrium}). As will become evident, our main result in this paper is a straightforward extension of Doob's result. 

Doob's fixed-time conditioning provides a framework for understanding how to relate a conditioned process to another process where the condition of the first is achieved automatically (i.e., without conditioning) at the price of applying an extra force. In Doob's original setting, the Markov process has a condition on an event at some fixed time, for example reaching a particular state. Among the many ways to realise a conditioning, the {\em Doob transform} \cite{chetrite2015nonequilibrium} is unique as it is optimal, in that it introduces 
only the minimal necessary perturbation to the dynamics of the original process, and the 
ensemble of trajectories generated by the new process is equivalent to the original trajectories post-selected to satisfy the condition. 

Consider a general (hard) condition on the state of the Markov chain at the final time $T$,  
\begin{equation}
\label{eq:DoobConditioningObservable}
    X_T \in D \, , \text{with} \; D \subset \Xi
    \, . 
\end{equation}
If the probability of trajectory $\omega_{0:T}$ is ${\mathbb P}[\omega_{0:T}]$, that of the conditioned process reads
\begin{equation}
    \label{eq:PathProbaOriginalDoob}
    {\mathbb P}[\omega_{0:T}] \, {\mathbb 1}[X_T \in D]
    \, , 
\end{equation}
where we denote by ${\mathbb 1}[\cdot]$ the indicator function, which is one if $(\cdot)$ is true, and zero otherwise (also called a ``projector''). \Er{eq:PathProbaOriginalDoob} is expressed as ``post-selection'': one imagines generating first the trajectory 
$\omega_{0:T}$ through the Markov dynamics, and then keeping it or discarding it depending on whether its endpoint satisfies \er{eq:DoobConditioningObservable}. Furthermore, note that if one wanted to define the ensemble of conditioned trajectories, then \er{eq:PathProbaOriginalDoob} also requires normalisation. Clearly, computing the conditioned ensemble in this way is highly inefficient (and, in practice, impossible in many cases). 

The probability of trajectories satisfying the condition can also be written as
\begin{equation}
    \label{eq:PathProbaOriginalDoob2}
    {\mathbb P}[\omega_{0:T-1}, X_T \in D] = 
    {\mathbb P}[X_T \in D \, | \, \omega_{0:T-1}] {\mathbb P}[\omega_{0:T-1}]
    =
    V_{T-1}(x_{T-1}) {\mathbb P}[\omega_{0:T-1}]
    \, ,
\end{equation}
where in the first equality we used Bayes theorem, and in the second we used the Markov property and defined 
\begin{equation}
    V_{T-1}(x) 
    \coloneqq
    {\mathbb P}[X_T \in D \, | \, X_{T-1} = x] 
    \, .
\end{equation}
We can define similar functions at other times for the probability of reaching the target condition at $T$ starting at state $x$ at time $t$, 
\begin{equation}
    \label{eq:SpaceTimeHarmonic}
    V_{t}(x) 
    \coloneqq
    {\mathbb P}[X_T \in D \, | \, X_t = x] 
    \, . 
\end{equation}
These functions satisfy a backward propagation equation
\begin{equation}
\label{eq:BKEDoob}
    V_{t-1}(x) = \sum_{y \in \Xi} p(x,y) V_t(y) \, ,
\end{equation}
with final condition 
\begin{equation}
\label{eq:FinalConditionDoob}
    V_T(x) = {\mathbb 1}[x \in D]
    \, .
\end{equation}
In analogy with Markov decision processes \cite{sutton2018reinforcement} we call $\{ V_t(\cdot) \}_{0:T}$ {\em value functions} 
(as they encode the ``value'' of states in terms of the likelihood of trajectories emanating from them to reach the target). \Er{eq:BKEDoob} propagates backward in time the information (or correlations) of the condition happening at the final time $T$. Using the value functions it is possible to derive a time-inhomogeneous Markov process $(\tilde{X}_t)_{0 \leq t \leq T}$, which we call the {\em Doob process}, which automatically satisfies the conditioning \eqref{eq:DoobConditioningObservable}. Using the value functions, the Doob dynamics is given by time-dependent transition probabilities $\left\lbrace \tilde{p}_t(x,y) \right\rbrace_{x,y=1:d}$, which read
\begin{equation}
\label{eq:TransitionMatrixDoob}
\tilde{p}_t(x,y) 
    = 
    \frac{ V_t(y)}{ V_{t-1}(x)} p(x,y) 
    \, , 
\end{equation}
where \er{eq:BKEDoob} guarantees the normalisation $\sum_y \tilde{p}_t(x,y) = 1$ for all $x$ and $t$. The corresponding path probabilities are
\begin{equation}
\label{eq:PathProbaDoob}
\begin{split}
    \mathbb{P}[\tilde{X}_0 = x_0, \cdots, \tilde{X}_T = x_T]
    &= 
    \tilde{p}_0(x_0) 
    \prod_{t=1}^T \tilde{p}_t(x_{t-1}, x_t) 
    = 
    \tilde{p}_0(x_0) 
    \prod_{t=1}^T \frac{ V_t(x_t)}{ V_{t-1}(x_{t-1})} p(x_{t-1},x_t) 
    \\
    &=
        \frac{\tilde{p}_0(x_0)}{V_0(x_0)} 
        P[X_1 = x_1, \cdots, X_{T-1} = x_{T-1}, X_T = x_T \, | \, X_0 = x_0] \, {\mathbb 1}[x_T \in D]
    \, ,
\end{split}
\end{equation}
where in the last equality we have used the telescopic nature of the product and \er{eq:FinalConditionDoob}. We note the following: (i) \er{eq:PathProbaDoob} guarantees that all trajectories of the Doob dynamics satisfy the condition \eqref{eq:DoobConditioningObservable}; (ii) if we choose $\tilde{p}_0 = p_0$ then \er{eq:PathProbaDoob} is the normalised conditioned probability 
\eqref{eq:PathProbaOriginalDoob}; (iii) $V_0$ is the normalisation of \er{eq:PathProbaOriginalDoob}, and plays the role of a dynamical partition function for the conditioned ensemble (and whose logarithm is the ``free-energy'' cost of transforming the original trajectory ensemble into the conditioned one). The above shows that by applying the Doob ``force'', cf.\ \er{eq:TransitionMatrixDoob}, we obtain a stochastic process that optimally generates the subensemble of conditioned trajectories of the original dynamics.

\subsection{Multi-time Doob conditioning on the empirical occupations}

The previous result can be extended to the case of conditioning through observables that depend on the trajectory at multiple times. We are in particular interested in (hard and/or soft) conditions like \era{eq:Conditioning}{eq:ConditioningValues} on the 
empirical occupation measure of the original Markov chain $(X_t)_{0 \leq t \leq T}$ at all times in the trajectory. 

We consider first the case of a hard constraint, cf.\ Sec.\,III.A.1. We proceed through analogous steps to those in Sec.\,IV.A above. We start by writing the probability of a trajectory $\omega_{0:T}$ satisfying condition \eqref{eq:ConditioningValues} in a post-selection form
\begin{equation}
    \label{eq:PathProbability}
    \begin{split}
        {\mathbb P}[\omega_{0:T}] \, 
            {\mathbb 1}[\mathbf{F}_T(\omega_{0:T}) \in \mathbf{D}_T]
        &= 
        \mathbb{P}[X_0 = x_0, \cdots, X_T = x_T] \, 
        \prod_{t=1}^T    
        {\mathbb 1}\left[ f_t(\rho_t(\omega_{0:t})) \in D_t \right]
        \\
        &=
        p_0(x_0) \prod_{t=1}^{T} p(x_{n-1},x_n) 
        {\mathbb 1}\left[ f_t(\rho_t(\omega_{0:t})) \in D_t \right]
        \, . 
    \end{split}
\end{equation}
Similar to \er{eq:DoobConditioningObservable} for the case of single-time conditioning, \er{eq:PathProbability} is not normalised over the set of conditioned trajectories. Notice that in the bottom line of \er{eq:PathProbability} the trajectory constraint is sliced into local-in-time constraints. 

We define a set of value functions that generalise \er{eq:SpaceTimeHarmonic}, that is, give the probability of satisfying the constraints in the future starting from the current conditions,
\begin{equation}
    \label{eq:Vs}
    V_{t}(x,\rho_t) 
    \coloneqq
    {\mathbb P}[ f_{t+1}(\rho_{t+1}) \in D_{t+1}, \cdots, f_{T}(\rho_{T}) \in D_{T} 
    \,        
    | \, X_t = x, \rho_t] 
    \, . 
\end{equation}
Notice that in contrast to \er{eq:SpaceTimeHarmonic}, the value functions defined in \er{eq:Vs} depend both on the state {\em and} the empirical occupation at time $t$. They obey the following backward equation
\begin{equation}
    \label{eq:BackwardGeneral}
    V_{t-1}(x, \rho_{t-1}) = \sum_{y \in \Xi} p(x,y) 
    \, {\mathbb1}[f_t(\rho_t^{(y)}) \in D_t]
    \, V_t(y, \rho_t^{(y)})
    \, , 
\end{equation}
where $\rho_t^{(y)}$ is the empirical that follows from $\rho_{t-1}$ under the $x \to y$ transition, $\rho_t^{(y)}(\cdot) = [t \rho_{t-1}(\cdot) + \delta_{(\cdot),y}]/(t+1)$. Since the projector appears explicitly in \er{eq:BackwardGeneral}, the final condition for the backward dynamics is
\begin{equation}
    \label{eq:FinalGeneral}
    V_{T}(x, \rho_{T}) = {\mathbb1}[f_T(\rho_T) \in D_T]
    \, .
\end{equation}

We can now define the transition probabilities of the Doob process,
\begin{equation}
    \label{eq:TransitionMatrixDoobGeneral}
    \tilde{p}_{t}(x,y) \coloneqq 
    \frac{V_{t}(y, \rho_t^{(y)})}{V_{t-1}(x, \rho_{t-1})}
    p(x,y) \, {\mathbb1}[f_t(\rho_t^{(y)}) \in D_t]
    \, .
\end{equation}
\Er{eq:BackwardGeneral} guarantees that these transition probabilities are normalised. In contrast to the case of one time conditioning, Sec.~IV.A, the transition probabilities \eqref{eq:TransitionMatrixDoobGeneral} depend on the empirical, and thus define a non-Markovian process. The probability of a trajectory under this dynamics is
\begin{equation}
    \label{eq:PathProbaDoobGeneral}
    \begin{split}
        \mathbb{P}[\tilde{X}_0 = x_0, \cdots, \tilde{X}_T = x_T]
        &= 
        p_0(x_0) 
        \prod_{t=1}^T \tilde{p}_t(x_{t-1}, x_t) 
        = 
        p_0(x_0) 
        \prod_{t=1}^T 
        \frac{V_{t}(y, \rho_t^{(y)})}{V_{t-1}(x, \rho_{t-1})}
        p(x,y) \, {\mathbb1}[f_t(\rho_t^{(y)}) \in D_t]
        \\
        &=
            \frac{p_0(x_0)}{V_0(x_0)} 
            P[X_1 = x_1, \cdots, X_{T-1} = x_{T-1}, X_T = x_T \, | \, X_0 = x_0] \,  {\mathbb 1}[\mathbf{F}_T \in \mathbf{D}_T]
        \, ,
    \end{split}
\end{equation}  
where we have set the initial probability of the Doob process to be the same as that of the original process, $\tilde{p}_0 = p_0$, and we write the value function at the start, $V_0(x_0, \rho_0) = V_0(x_0)$, since $\rho_0(x) = \delta_{x,x_0}$. 

\Er{eq:PathProbaDoobGeneral} shows that the Doob dynamics directly generates the multi-time conditioned ensemble. However, the Doob transition probabilities \eqref{eq:TransitionMatrixDoobGeneral} are of the self-interacting form \eqref{eq:TransitionSIMC}. This is the central result of the paper, the proof that the optimal sampling dynamics of a Markov process subject to conditioning via functions of its empirical occupations is given by a non-Markovian SIP. 

While we proved this result for hard conditioning, all the steps above carry through if the projectors ${\mathbb 1}[\cdot]$ are replaced by (possibly time dependent) weight functions $W[\cdot]$ that implement a soft conditioning. The examples in Sec.~V illustrate both kinds of conditionings.

\begin{figure*}[t]
    \centering
    \includegraphics[width=0.95\textwidth]{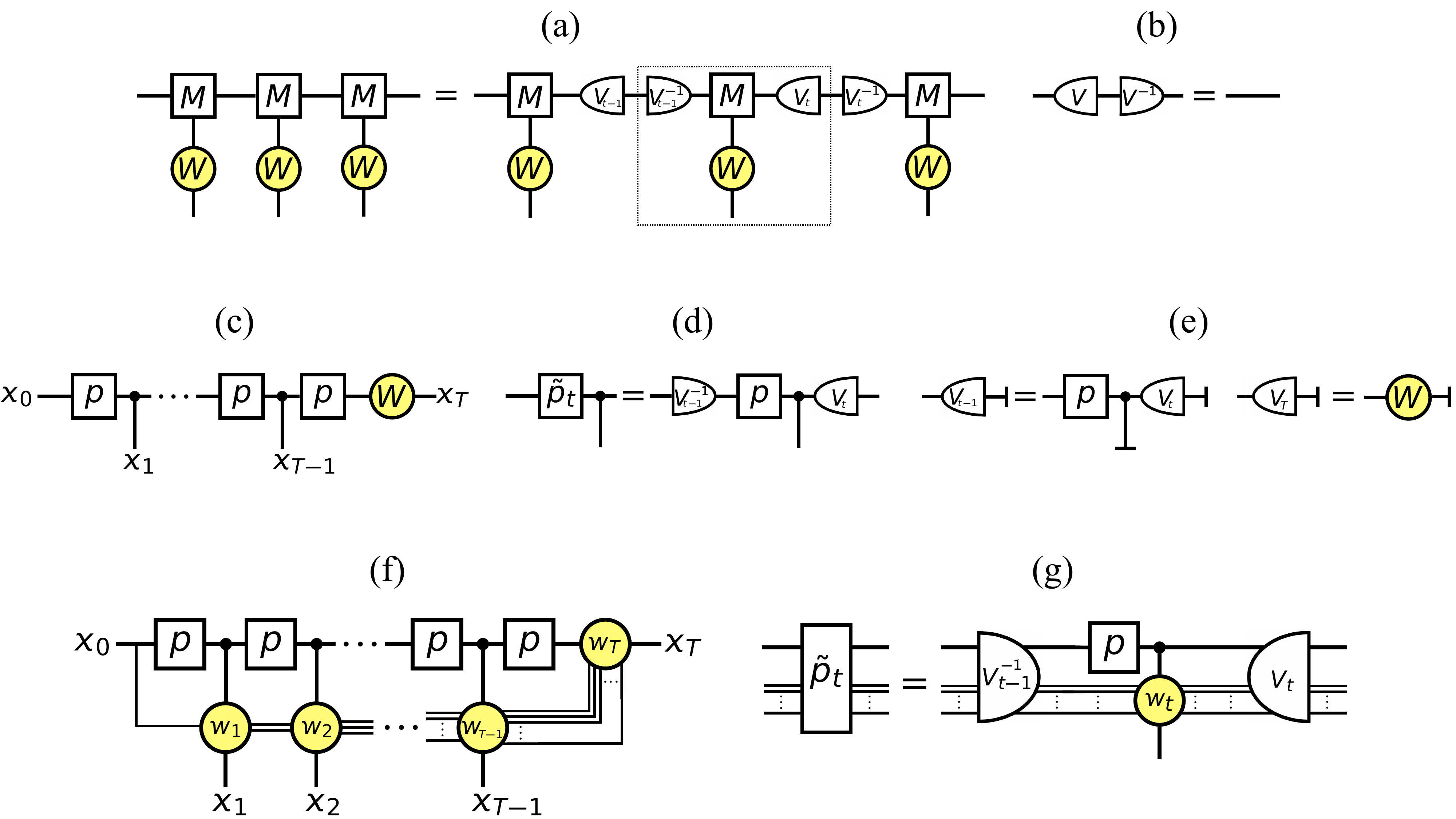}
    \caption{
    {\bf Local-in-time conditioning.} 
        {\bf (a)} The l.h.s.\ represents a conditioned dynamics. The rank-2 tensors $W$ (yellow circles) are diagonal operators that impose conditions at each time by multiplying the probability of a trajectory by a factor that depends on the observed transition at that time. 
        For hard conditioning, cf.\ Sec.\,III.A.1, the weighing functions are projectors taking values 0 or 1, while for soft conditioning, cf.\ Sec.\,III.A.2, they introduce non-negative factors.        
        The overall reweighting of the trajectory probability is the product of all these factors. In the r.h.s.\ we have inserted identities. We see that the same TN is obtained if we define the elementary tensors in terms of the operators collected inside the dotted box. This illustrates the ``gauge symmetry'' of the TN which we can exploit to obtain the Doob dynamics of the conditioned process. Specifically, we aim to find {\em value functions} $V_t$ such that the operators in the box are in right canonical form. This corresponds to {\em fixing the gauge} of the TN. 
        {\bf (b)} Identity used in the r.h.s.\ of the previous panel in terms of diagonal rank-2 operators corresponding to the value functions. 
    {\bf Basic Doob transform.} 
        {\bf (c)} TN of a Markov process with a condition on its final state. 
        {\bf (d)} Exploiting the gauge symmetry of the TN, see panel (a), we obtain the transition operators of the Doob process that optimally generates conditioned trajectories, see \er{eq:TransitionMatrixDoob}. 
        {\bf (e)} Graphical representation of \era{eq:BKEDoob}{eq:FinalConditionDoob}. Value functions that satisfy these equations make the tensors in the previous panel to be of right canonical form. 
    {\bf Multi-time conditioning and generalised Doob transform.} 
        {\bf (f)} TN representing conditioning dependent on the states visited in the past. The conditioning operators, $w_t$, are diagonal tensors that receive information on the past states in the trajectory and also pass it forward.     
        {\bf (g)} The corresponding Doob dynamics, \er{eq:TransitionMatrixDoobGeneral}, is Markovian only in the extended space $\Xi \times \Gamma$. 
    }  
    \label{fig:fig2}
\end{figure*}

\subsection{Tensor network description}

The results of Secs\,IV.A and IV.B can be easily understood using TN concepts. The l.h.s.\ of Fig.\,2(a) shows a local-in-time conditioning/biasing. One can insert identities between the transitions given by the product of a value function and its inverse, cf.\ Fig.\,2(b). This gives the r.h.s.\ of Fig.\,2(a). By grouping the local tensors as shown by the dotted box in Fig.\,2(a) we see that this corresponds to a {\em gauge transformation} of the TN \cite{verstraete2008matrix,banuls2023tensor}. The Doob transform amounts to finding suitable value functions such that these transformed local tensors are those of a probability conserving dynamics. This {\em gauge fixing} \cite{garrahan2016classical} allows to absorb the local-in-time condition/bias into the (unconditioned) transition operators of a new (Doob) dynamics. Figure\,2(c-e) shows this procedure for the Doob conditioning results of Sec.\,IV.A: panel (c) is the TN for the dynamics conditioned at the final time; panel (d) is the graphical representation of \er{eq:TransitionMatrixDoob} (i.e., the gauge transformation of the TN); and panel (e) is the graphical representation of \er{eq:BKEDoob} (i.e., the gauge fixing).

As shown in Sec.\,IV.A, the Doob process is a Markovian dynamics in the same state space $\Xi$ as the of the original process. This can be understood from the corresponding TNs: for local-in-time conditioning, as in Figs.\,2(a) or 2(c), the MPO that is applied to the MPS that represents the trajectory ensemble is a {\em product operator} (an MPO of bond dimension one) \cite{verstraete2008matrix,banuls2023tensor}. As such, the combined TN has the same bond dimension as the original TN. This is seen by comparing the dimensions traversed by a vertical cut in Fig.\,1(c) vs 2(a), or Fig.\,1(g) vs 2(c). That is, the conditioning does not increase the information that needs to be propagated from left to right, or in other words, the conditioned process is as Markovian as the unconditioned one. 

The situation is very different for multi-time conditioning as in Sec.\,IV.B. Figure\,2(f) shows the TN describing a condition that depends on past occupations. The biasing operator is no longer a product operator but an MPO with the bond dimension required to receive and pass forward information about visited states. It is clear that a vertical cut of the conditioned Fig.\,2(f) traverses more dimensions that a vertical cut of Fig.\,1(g). For conditioning dependent on the empirical occupations, cf.\ \er{eq:PathProbability}, the bond dimension of Fig.\,2(f) becomes $\Xi \times \Gamma$ (that of the original process times that of the MPO that imposes the condition/bias). This means that in contrast to local-in-time conditioning, multi-time conditioning introduces non-Markovian effects for the dynamics in state space $\Xi$. The resulting generalised Doob transformation is shown in Fig.\,2(g), expressed as a Markov process in an extended state space, corresponding to a TN representation of \er{eq:TransitionMatrixDoobGeneral}.

\begin{figure*}[t]
    \centering
    \includegraphics[width=\textwidth]{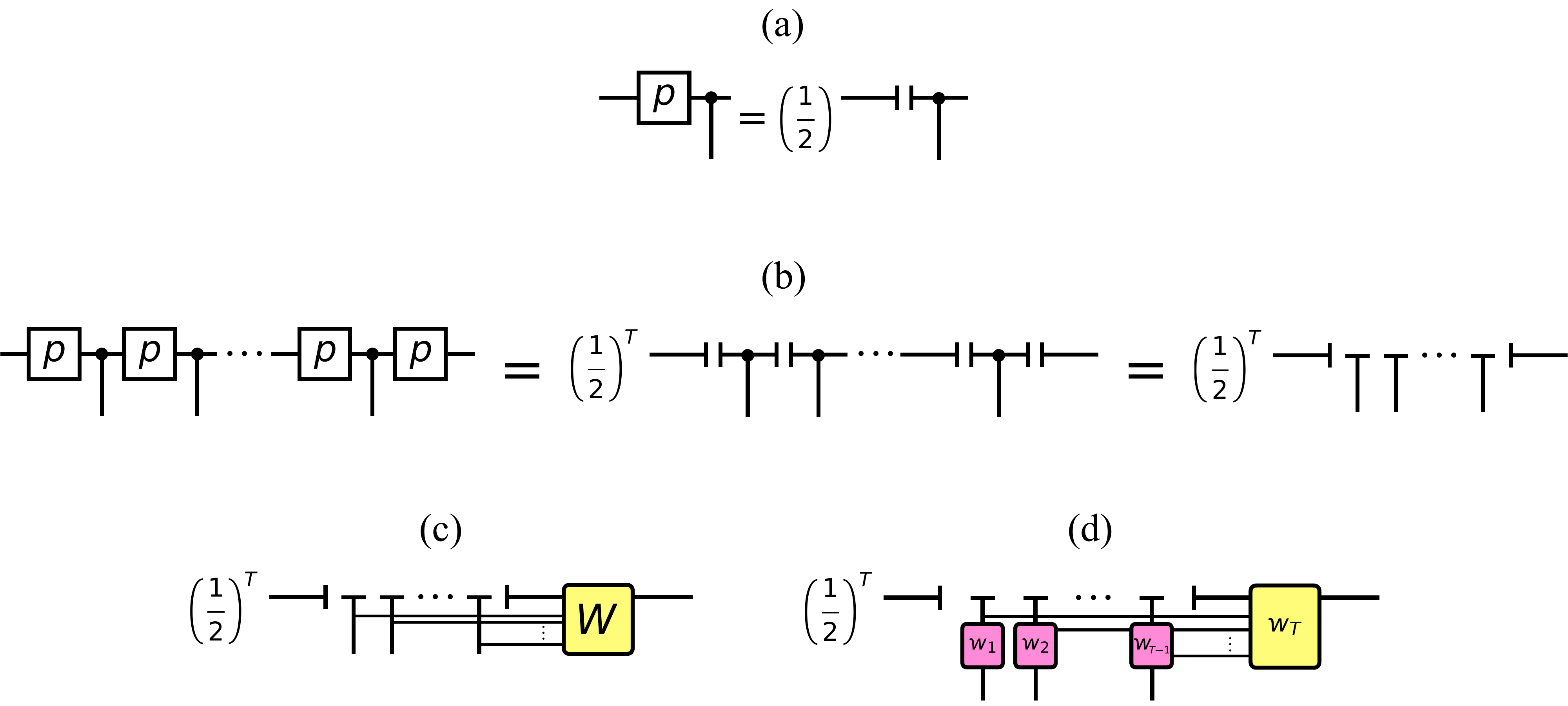}
    \caption{
        {\bf Fair coin and conditionings.} 
            {\bf (a)} The elementary tensor for the coin transitions, \er{eq:pCoin}, is the product of flat state. 
            {\bf (b)} The MPS representing the dynamics of the coin becomes a product operator as no information needs to be carried along the time direction, that is, every transition is independent of the previous one, and the Markov chain is an i.i.d.\ process. 
            {\bf (c)} TN for the conditioning of Example A, \er{eq:CondBridge}. {\bf (d)} TN for the conditionings of Examples B and C, \era{eq:CondExcursion}{eq:FavourHeight}.
    }  
    \label{fig:fig3}
\end{figure*}

\section{Examples}
\label{sec:examples}

We now illustrate our general results above with three elementary examples. For the unconditional dynamics we consider the simplest stochastic process, that of a fair coin. That is, the system has only two states, $\Xi = \{-1, 1 \}$, and the transitions between states in the unconditioned dynamics have the same probability, so that the transition matrix, cf.\ \er{eq:pxy}, reads
\begin{equation}
    \label{eq:pCoin}
    p
    =
    \frac{1}{2}
    \begin{pmatrix}
    1 & 1 \\
    1 & 1 
    \end{pmatrix} \, .
\end{equation}
Since the matrix above has rank one, the TN that describes the dynamics reduces to a product state (an MPS with bond dimension 1), meaning that no information has to be carried along the time direction; see Fig.\,3(a,b). 

Below we introduce three kinds of conditions and obtain the associated SIPs that realise them optimally. For all three cases the trajectory observable \eqref{eq:Conditioning} will have functions
\begin{equation}
    \label{eq:fRW}
    f_t(\rho_t(\cdot \, | \, \omega_{0:t})) 
    = 
    \rho_t(+1 \, | \, \omega_{0:t}) - \rho_t(-1 \, | \, \omega_{0:t})
    \, .
\end{equation}
If we interpret each flip of the coin as a step to the left or right on a one-dimensional lattice, each function $f_t$ gives the position after $t$ steps of an unbiased random walk (RW), 
\begin{equation}
    \label{eq:nt}
    n_t \coloneqq t \, f_t 
    \, .
\end{equation}
For trajectories up to time $T$ we have
\begin{equation}
    \label{eq:ntspace}
    n_t \in \Lambda \coloneqq \{ -T, -T + 1, \cdots, 1, 0, 1, \cdots, T-1, T \} 
    \, .
\end{equation}
Clearly, the space $\Lambda$ of the RW is isomorphic to that of the empirical occupation of the coin process, $\Gamma$.

\subsection{Random walk bridge}
\label{subsec:bridge}

We consider first a hard condition, cf.\ Sec.\,III.A.1, and for \er{eq:ConditioningValues} we choose
\begin{equation}
    \label{eq:CondBridge}
    \mathbf{D}_T =  \left( [-1,1], [-1,1],  \cdots, \{ 0 \} \right)  \, .
\end{equation}
Condition \eqref{eq:ConditioningValues} with \er{eq:CondBridge} is automatically satisfied for $t<T$ since $-1 \leq f_t \leq 1$ by definition. The only non-trivial condition is at the final time, $f_T = 0$. This conditions trajectories to have as many transitions to states $X=1$ as to $X=-1$, irrespective of the initial configuration (here and in what follows we assume $T$ is even). In terms of the location of the RW $n_t$, this corresponds to conditioning a walker that starts at the origin to return to the origin at the final time $T$, a process known as a {\em random walk bridge} \cite{majumdar2005airy,majumdar2006brownian,majumdar2015effective,causer2022slow,grimmett2020probability}.

The TN for the conditioned process is shown in Fig.\,3(c). It is a special instance of the multi-time conditioning, Fig.\,2(f): while the MPS of unconditioned dynamics of the coin, Fig.\,3(b), is a product state with bond dimension 1, the information that needs to be propagated for the final time condition \eqref{eq:CondBridge} increases the bond dimension in the TN of Fig.\,3(c). 

Given the multi-time nature of the condition, it is not possible to define the optimal sampling Doob dynamics as a Markov process in the space $\Xi$ of the coin alone, and we need to extend to the space of the empirical occupation. In order to obtain the generalised Doob dynamics we proceed as in Sec.\,4.B. The key quantities are the value functions \eqref{eq:Vs}. For the bridge condition, the value function at time $t$ can be obtained from the fraction of RW paths that start at $n_t$ and reach $n_T=0$. Simple combinatorics gives 
\begin{equation}
    \label{eq:VsBridge}
    V_{t}(x,n_t) 
    = 
    2^{T-t}
    {T - t \choose {\frac{1}{2}(T - t - n_t)}}
    \;\;\;  \text{for} \;
    |n_t| \leq T-t 
    \, , 
\end{equation}
where we replace the dependence on the empirical $\rho_t$ for that of the RW $n_t$ given the equivalence of $\Gamma$ and $\Lambda$, and $V_{t}(x,n_t)=0$ otherwise, as the position of the walker $n_t$ has to be within the ``light-cone'' emanating from the origin for it to be able to reach the condition within the remaining time $T-t$. Note that the value function is independent of the current state of the coin $x$ (a consequence of the i.i.d.\ nature of the unconditioned dynamics). It is easy to check that the value functions \eqref{eq:VsBridge} satisfy the backward propagation \er{eq:BackwardGeneral}. Inserting it into \er{eq:TransitionMatrixDoobGeneral} we get the $n_t$-dependent transition matrix for the Doob dynamics that generates the conditioned trajectories of the coin
\begin{equation}
    \label{eq:SelfInteractingBrownian}
    \tilde{p}_t = 
    \frac{1}{2}
    \begin{pmatrix}
        1
        \\
        1 
    \end{pmatrix} 
    \left(
    \begin{array}{ccc}
        1 - \frac{n_t}{T-t+1}
        & &
        1 + \frac{n_t}{T-t+1}
    \end{array} 
    \right)
    \, .
\end{equation}
We therefore see that a condition on the accumulated transitions of the coin being exactly balanced at the final time, \era{eq:fRW}{eq:CondBridge}, is optimally realised by the SIP process \er{eq:SelfInteractingBrownian}. Futhermore, this SIP in the coin space $\Xi$ is a Markovian process in the extended RW space $\Lambda$: if we denote by $\{ q(n,n') \}_{n,n' \in \Lambda}$ the transition probabilities of an unbiased RW, with 
\begin{equation}
    \label{eq:qRW}
    q(n,n') = \frac{1}{2} \delta_{n', n \pm 1}
    \, ,
\end{equation}
then \er{eq:SelfInteractingBrownian} becomes a RW with a time-dependent force, 
\begin{equation}
    \label{eq:SimplifiedDoobBrownianBridge}
    \tilde{q}_t(n, n \pm 1) 
    = \frac{1}{2} \left( 1 \mp \frac{n}{T - t + 1} \right) 
    \, .
\end{equation}
The dynamics \er{eq:SimplifiedDoobBrownianBridge} is precisely that of RW bridges for time $T$, see e.g.~\cite{causer2022slow}.

\subsection{Random walk excursion}
\label{subsec:excursion}

As a second example we consider again the fair coin with  a hard condition on observable \eqref{eq:fRW}, but which now is non-trivial for all times. Specifically, we set  
\begin{equation}
\label{eq:CondExcursion}
    \mathbf{D}_T = \left( [0,1], [0,1], \cdots, \{ 0 \} \right) \, . 
\end{equation}
The final time condition is the same as in Example A, requiring the overall number of flips of the coin to be exactly balanced at time $T$. The conditions for $t < T$, however, also impose the restriction that the number of down-flips cannot exceed the number of up-flips at any time. The TN for this condition is shown in Fig.\,3(d), the biasing functions $w_t$ being projectors, $w_t = {\mathbb 1}[f_t \in D_t]$, where $D_t$ are the sets of \er{eq:CondExcursion}. In terms of the RW position (i.e., of the empirical occupation), the condition forces $n_T=0$, with $n_t \geq 0$ at all $t<T$. That is, the multi-time condition restricts the RW paths to be bridges that do not cross the origin at any time, also known as {\em excursions} \cite{majumdar2005airy,majumdar2006brownian,majumdar2015effective,causer2022slow,grimmett2020probability}. 

As in the previous example, the value functions \eqref{eq:Vs} are obtained by counting the number of paths that reach the the origin at the final time, but now also having to stay non-negative throughout. While the combinatorics leading to \er{eq:VsBridge} for bridges comes from the Pascal triangle, 
that for the excursions comes from the Catalan triangle, 
\begin{equation}
    \label{eq:VsExcursion}
    V_{t}(x,n_t) 
    = 
    2^{T-t}
    \left[ 
    {T - t \choose {\frac{1}{2}(T - t - n_t)}}
    - 
    {T - t \choose {\frac{1}{2}(T - t - n_t)-1}}
    \right]
    \, .
\end{equation}
In this case, the Doob process for the coin is given by 
\begin{equation}
    \label{eq:SelfInteractingExcursion}
    \tilde{p}_t = 
    \frac{1}{2}
    \begin{pmatrix}
        1
        \\
        1 
    \end{pmatrix} 
    \left[
    \begin{array}{ccc}
        \left( 1 + \frac{1}{n_t+1} \right)
        \times
        \left( 1 - \frac{n_t}{T-t+1} \right)
        & & 
        \left( 1 - \frac{1}{n_t+1} \right)
        \times
        \left( 1 + \frac{n_t+2}{T-t+1} \right)
    \end{array} 
    \right]
    \, ,
\end{equation}
and we see that the optimal dynamics to sample the conditioned coin process is also a SIP. 

The Doob dynamics in the extended state space $\Lambda$ of the RW in this case reads
\begin{equation}
    \label{eq:SimplifiedDoobBrownianExcursion}
    \tilde{q}_t(n, n \pm 1) 
    = \frac{1}{2} 
    \left( 1 \pm \frac{1}{n+1} \right)
    \left( 1 \mp \frac{n + 1 \mp 1}{T-t+1} \right)
    \, ,
\end{equation}
which implements the forces on a RW so that the trajectories are excursions.

\begin{figure*}[t]
    \centering
    \includegraphics[width=\textwidth]{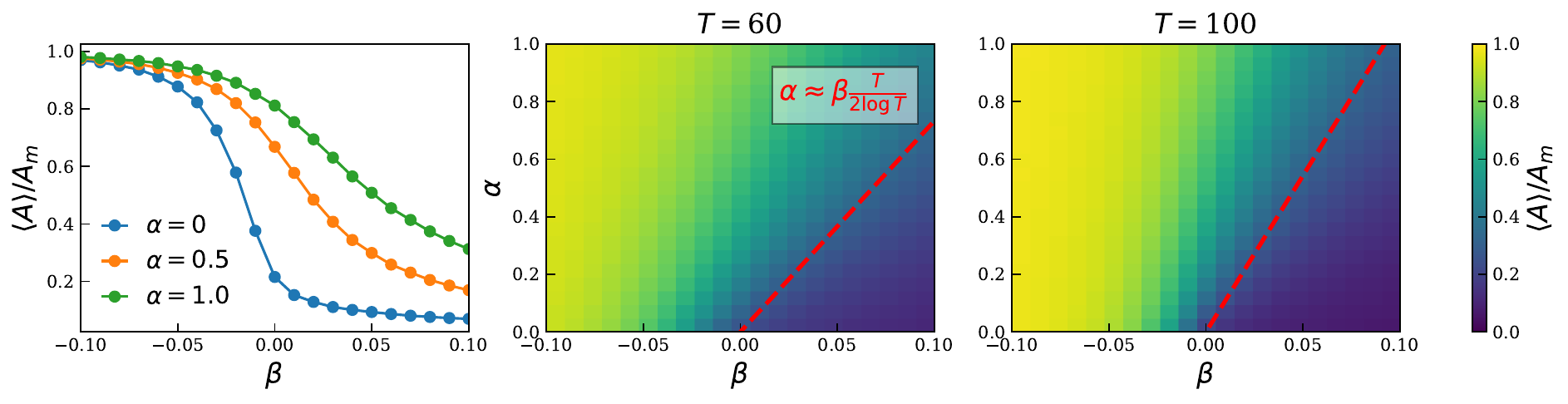}
    \caption{
        {\bf Dynamical phase diagram of the Example C.} 
        We compute the order parameter \er{eq:area} by sampling the conditioned trajectory ensemble with the optimal Doob dynamics: we obtain the value functions by solving \er{eq:BackwardC} recursively in a numerically exact way (which limits the largest $T$ that we can simulate); from the value functions we obtain the transition probabilities \er{eq:C} for all configurations and times, which we use to run the dynamics. 
        {\bf (a)} Average area (scaled by the maximum area $T^2/4$) as a function of $\beta$ for three values of $\alpha$, at $T=100$. 
        {\bf (b)} 
        Average area as a function of $\alpha$ and $\beta$ for $T=60$. The dashed-red line shows the estimate for the crossover between the large and small area regimes. 
        {\bf (c)} Same for $T=100$. 
    }  
    \label{fig:fig3}
\end{figure*}

\subsection{Random walk excursion with competing forces}
\label{subsec:bridgecompeting}

As a final example we consider the same hard conditioning that leads to the RW excursions above, but with an extra trajectory biasing (soft conditioning) in terms of two competing factors dependent on the empirical occupation. This combination is defined in terms of weigthing functions that depend on the observables \eqref{eq:fRW},
\begin{align}
\label{eq:FavourHeight}
    w_t = {\mathbb 1}[f_t \in D_t]
    \, (n_t+1)^\alpha \, e^{-\beta n_t} 
    \, ,
\end{align}
where $D_t$ are the sets of \er{eq:CondExcursion}, $\alpha \geq 0$ and $\beta$ real and arbitrary. The projector in the first factor imposes the excursion constraint. The second factor is a polynomial cost from having a balanced coin empirical occupation (i.e., a repulsion from the origin of the RW). The third factor in contrast is an exponential cost for a non-balanced coin empirical occupation (i.e., an attraction of the RW towards the origin) if $\beta > 0$ (and a repulsion of the RW from the origin for $\beta <0$).  The introduction of these extra weights creates additional correlations in the RW excursions of Sec.\,V.B. 

The recursion relation \eqref{eq:BackwardGeneral} for the value functions  simplifies to 
\begin{equation}
    \label{eq:BackwardC}
    V_{t-1}(n_{t-1}) 
    =
    \frac{1}{2} 
    V_{t}(n_{t}+1)
    \theta(n_{t}+1)
    (n_t+1)^\alpha e^{-\beta (n_t+1)}
    +
    \frac{1}{2} 
    V_{t}(n_{t}-1)
    \theta(n_{t}-1)
    (n_t-1)^\alpha e^{-\beta (n_t-1)}
    \, , 
\end{equation}
with $V_T(n_T) = \delta_{n_T,0}$ and where $\theta(\cdot)$ is the Heaviside step function. Additionally, we have dropped 
the coin state $x$ from the argument of the value function as there is no dependence on it, cf.\ \era{eq:VsBridge}{eq:VsExcursion}. However, in contrast to the two examples above, for this case there is no simple closed form for the solution of \er{eq:BackwardC}. 

As before, the Doob dynamics that realises the conditioned process is a SIP in the configuration space of the coin $\Xi$, with transition probabilities of  \er{eq:TransitionMatrixDoobGeneral} using the value functions \eqref{eq:BackwardC}. Alternatively, the Doob dynamics can be described as a Markovian process in the RW space $\Lambda$, with rates 
\begin{equation}
    \label{eq:C}
    \tilde{q}_t(n, n \pm 1) 
    = \frac{1}{2}     
    \, (n + 1)^\alpha \, e^{-\beta n} 
    \frac{V_{t}(n \pm 1)}{V_{t-1}(n)}
    \, .
\end{equation}

The competition between the algebraic and exponential biasing factors in \er{eq:FavourHeight}, when $\alpha,\beta>0$, favouring large and small excursions, respectively, can give rise to a transition in the dynamics. A convenient 
order parameter is the area under the RW averaged over the dynamics,
\begin{equation}
\label{eq:area}
    \langle A \rangle 
    = 
    \left< \sum_{t=1}^{T} n_t \right>
    \, , 
\end{equation}
where $\left\langle \cdot \right\rangle$ marks the ensemble average. $\langle A \rangle$ takes values between a minimum of $T/2$ and a maximum of $A_m \coloneqq T^2/4$. This difference in scaling with $T$ suggests two possible regimes for large $T$. 

The case with $\alpha = 0$ was studied in the context of Fredkin spin chains in Ref.\,\cite{causer2022slow}, where the equilibrium properties of the one-dimensional Fredkin model map to the dynamical properties of a conditioned coin with Doob dynamics corresponding to \er{eq:C}. Using this equivalence we know that for $\alpha=0$ there is a phase transition \cite{causer2022slow} at $\beta=0$ in the limit of long time $T$, between a phase with $\langle A \rangle \propto T$ for $\beta>0$, to a phase with $\langle A \rangle \propto T^2$ for $\beta<0$. 

For finite $T$ the transition smoothes to a crossover occurring around $\beta = 0$. Since for $\alpha > 0$ large area trajectories are favoured, we expect that the crossover will move to positive $\beta$, increasingly so with increasing $\alpha$ (and becoming sharper for larger $T$). That is, when $\alpha$ is positive, a larger $\beta$ is needed to suppress the large area trajectories in favour of the small area regime. In Fig.\,4 we show results which give evidence for this behaviour (see caption for details of the numerics). Figure\,4(a) shows the average rescaled area as a function of $\beta$ for three values of $\alpha$: for $\alpha=0$ we recover the result of Ref.\,\cite{causer2022slow}, corresponding to a crossover that will sharpen into a phase transition for large $T$; for $\alpha>0$, as $\alpha$ increases the area grows for fixed $\beta$, as expected. Figures\,4(b,c) show the order parameter as a function of $\alpha$ and $\beta$ for two values of $T$: we see that there are two regimes, the large area one for small $\beta$ and large $\alpha$, and the small area one for large $\beta$ and small $\alpha$. The change appears to become sharper with larger $T$. We can estimate the location of the crossover from a simple heuristic argument, by equating the probability of a maximum area excursion with that of a minimum area one. This gives that for large $T$ the location of the transition should obey $\alpha \approx T \beta / (2 \ln T)$. The dashed-red line in Figs.\,4(b,c) shows that this is a reasonable estimate.

\section{Conclusion}
\label{sec:conclusion}

We have shown here that non-Markovian self-interacting processes, where transition probabilities depend on empirical occupations along a trajectory, in general correspond to the optimal sampling (or Doob) dynamics of a conditioned Markovian process, where the condition is non-local in time. We have also shown that this connection is made evident by representing trajectory ensembles in terms of tensor networks. We illustrated our general results with simple examples where conditioning the i.i.d.\ dynamics of a fair coin leads to self-interacting dynamics that in an extended space are random walks subject to time-dependent forces. We can anticipate several 
directions to consider following from these results. 
One is the study of other forms of non-Markovian dynamics and their relation to conditioned Markov processes. There should also be interesting connections to reinforcement learning problems with multi-time reward objectives. And another fruitful avenue of work should be the generalisation of the ideas here to open quantum dynamics.

\acknowledgements

We acknowledge financial support from EPSRC grant no.~EP/V031201/1.

\bibliography{bibliography-03032025}

\end{document}